\documentclass[12pt]{revtex4}
\begin{document}

\title{\Large \bf Quantum Theory and the Nature of Gravitation\footnote{Published in \textbf{``Quantum Leap''},
\textbf{Fall 2004 Issue}, edited by Dr. Ralf W. Gothe and Mrs.
Mary C. Papp. \textbf{``Quantum Leap''} is the bi-annual
publication of the Department of Physics and Astronomy, University
of South Carolina.}}
\author{Pawel O. Mazur}
\address{\sl Department of Physics and Astronomy \\
\sl University of South Carolina \\
\sl Columbia, S.C. 29208, U.S.A. \\
{\tt mazur@mail.psc.sc.edu} \\ {\tt mazur@physics.sc.edu}
\\ \\ \\ \\ \\
\qquad\qquad\qquad\qquad\qquad\qquad\qquad\qquad\qquad\qquad\qquad
{Dedicated to Professor Andrzej Staruszkiewicz on his birthday
\\ \\ \\ \\}}
\date{September 2004}
\begin{abstract} This is an essay sketching the line of thinking which
has led the present author to propose the constituent or atomic
model of gravitation more than a decade ago. It turns out that
viewing the problem of gravitation as a quantum many body problem
could be quite useful when addressing some old unsolved problems
such as the cosmological constant problem. I have applied this
idea in 1996 to the problem of the largest cold gravitating
system, the finite Universe itself. The result was the prediction
of a small, positive vacuum energy density, now known, after its
experimental discovery in 1998, as `dark energy'. The smallness of
this quantity was understood as the finite size effect in the cold
quantum many body system, and I quote here from~\cite{Ma96}, {\it
`` The smallness of the cosmological constant in natural Planck
units is a result of an almost perfect thermodynamical limit. This
is to say that the smallness of the cosmological constant is an
effect due to an enormous number $N$ of hypothetical
\textsf{gravitational} \textsf{atoms}. The present upper bound on
the cosmological constant $\Lambda$ allows us to draw the
conclusion about the lower bound on a number of
\textsf{gravitational} \textsf{atoms} in the observed Universe,
$N\sim 10^{122}$. The large numbers' coincidences noticed long
time ago have something in it after all. We have only one
reservation to add: elementary particles are not the same as the
hypothetical \textsf{gravitational} \textsf{atoms}, and,
therefore, have little to do with the cosmological constant. The
existence of the latter must be inferred indirectly from phenomena
as it was previously done for atoms and elementary particles. We
find an unusual coordination between the gravitational atomistic
aspect of physical reality in the regime usually called infrared,
or large distance scale, and the regime usually called
ultraviolet, or short distance scale.''} The old cosmological
constant problem is a man-made problem because the vacuum energy
density has nothing to do with the quartic divergences (zero point
energies) in the interacting relativistic quantum field theories
(i.e. in the Standard Model of elementary particles). The actual
value of the vacuum energy density of the vacuum finite size de
Sitter Universe is $\frac{E_{0}}{V}$, where $E_{0}$ is the ground
state energy of $N$ \textsf{gravitational} \textsf{atoms}, which
are spin-zero bosons of mass $M\sim M_{_{Pl}}$, contained in the
finite volume $V$.

\end{abstract}
\maketitle

\newpage

I wish to describe to you the line of thought which led me to the
conclusion that gravitation may, in fact, be the most spectacular
demonstration of quantum phenomena on the macroscopic scale.

\vskip .1cm \noindent What is the central idea of this direction
of research?

\noindent
The central idea is that gravitational
phenomena are best understood as quantum many body problem. The
microscopic theory of gravitation should be based on the idea of
{\it gravitational constituents}. Gravitational constituents are
heavy bosons. The bosonic nature of these constituents follows
from the nature of gravitation perceived as a macroscopic quantum
phenomenon. The appellation `heavy' refers to the extreme
feebleness of gravitational forces. A large number of heavy bosons
in three spatial dimensions can behave macroscopically in a
classical way so that the concept of gravitational field in the
large can be defined. The situation here is not unlike the one we
have encountered previously in the development of theoretical
physics.

\vskip .1cm \noindent Why is the concept of macroscopic
electromagnetic fields applicable to quantum states of a large
number of photons?

\vskip .1cm \noindent This is because photons are bosons.

\vskip .2cm \noindent Why is this way of looking at gravitation
useful and what reasons do we have to think that the constituent
picture may be the correct one?

\vskip .1cm \noindent The rationale here is the prediction and
explanation of new phenomena some of which have been discovered
relatively recently.

\vskip .1cm \noindent Reading Einstein's {\it The Meaning of
Relativity} one may come to the conclusion that gravitation and
hydrodynamics have much in common. The question then was:

\noindent What kind of hydrodynamics?

\vskip .1cm \noindent Regarding gravitation as a kind of
hydrodynamics led me to look for vortices in the gravitational
setting of general relativity~\cite{Ma86,Ma87}. My thinking about
gravitation in these terms was also strongly influenced by some
work on two-dimensional gravitation~\cite{AS63} which my PhD
Thesis advisor Andrzej Staruszkiewicz has done in the early 60s.
It turns out that there are no perturbative excitations
(gravitons) in the lower dimensional gravitation. However, one
finds that there are non-perturbative excitations such as
vortices. This essential difference between Einsteinian
gravitation in three spatial dimensions and the Staruszkiewicz
gravitation~\cite{AS63} on the two-dimensional plane, regarded as
an idealization of the thin surface layer in three-dimensional
space, has alerted me very early on to the possibility that
gravitational phenomena may have something to do with properties
of some peculiar medium. I started to look for examples of
`atomic' media where one sees such dramatic dependence on the
dimensionality of space. It is known that a system of heavy bosons
interacting via short range forces may undergo Einstein
condensation in three spatial dimensions but it cannot condense at
even very small temperatures in two spatial dimensions. But how
can one show that gravitation has something to do with the
presence of heavy bosonic constituents? This is here where the
study of thermal properties of quantum black holes helps. I have
found out that thermodynamical properties of black holes and
cosmological spaces with event horizons indeed do suggest the
existence of heavy constituents responsible for the feebleness of
gravitational forces. It turned out that the low temperature
thermal properties of quantum black holes are dominated by the low
energy long wavelength collective excitations that behave like
gapless phonons. However, this in itself does not imply that the
heavy constituents must be fundamental bosons. One can easily
envisage the situation where one has composite bosons such as the
Cooper pairs in the p- or the d-wave superconductors or in the
superfluid He3. The isotropy of the velocity of light puts severe
constraints on such scenario. The medium must be isotropic as far
as the propagation of light and gravitons is concerned. This
follows from the unusually precise tests of special relativity.
Hence, one may conclude that there is no emergent gravitation in
superfluid phases of He3, contrary to some statements one
encounters in the literature.

The origins of the attempt at the microscopic theory
of gravitation go back to my studies of gravitational
vortices~\cite{Ma86,Ma87}, perceived as analogs of Abrikosov
vortices in superconductors rather than analogs of the
Onsager-Feynman vortices in superfluid He4~\cite{ChapMa04}, and
the seemingly separate from it problems of gravitational collapse
of matter and the non-vanishing positive vacuum energy
density~\cite{Ma96}. The latter problem is also known as the
cosmological constant problem. The distinction between the two
kinds of vortices that may serve as analogs of the gravitational
vortex becomes clear-cut once one takes into account the basic
fact of experience that there exist inertial frame dragging
forces. Already in his first papers on the relativistic theory of
gravitation Einstein has asked the following question: Is there an
analog of magnetic fields in gravitation? These days we have the
name for it: gravimagnetic fields. In fact, the Gravity Probe B
mission, launched belatedly in April 2004, is dedicated to the
detection of the gravimagnetic field of the Earth. We now know
about their presence in the vicinity of pulsars (neutron stars).

\vskip .1cm If gravitation has something to do with Einstein
condensation of bosons then we should single out the best
theoretical laboratory where this hypothesis can be tested. It
turns out to be the gravitational vortex which I have discovered
long time ago. The breakdown of quantum fluid rigidity (the
breakdown of the Einstein condensate of bosons) should then lead
to the easily seen pathologies in the general relativity
description of physical situations. One does not need to look too
far. It is known that general relativity fails in the regions of
extremely high tidal forces of the type of a Big Bang or the
interior of analytically continued black hole solutions of general
relativity. It also fails on the event horizons of black holes and
on the cosmological event horizons. A third kind of the failure of
general relativity (also on the macroscopic length scales) is
associated with the occurrence of closed time-like curves (CTC) or
`time machines'. CTCs occur frequently in analytically extended
space-times described by general relativity once there is rotation
present in a physical system under consideration, which is quite
common in nature.

\vskip .1cm These three, seemingly separate until relatively
recently, instances of the failure of general relativity are now
seen as intrinsically connected. One should always look for other
instances of breakdown of a given theory after one finds out that
this theory breaks down in one instance. From this point of view
it is clear that black hole event horizons and cosmological
horizons were to be seen as the natural candidates of instances of
such a breakdown of the theory. But this realization has been long
in coming. I have reached this conclusion quite early but the
actual demonstration of it in technical terms came only in the
90s~\cite{Ma95,Ma96,Ma97a,Ma97b,Ma97c,MaMot98}. The later work on
quasi black hole objects (QBHOs or gravastars) has brought the
point in to the broader scientific
community~\cite{MaMot01,MaMot04}.

\vskip .1cm It is remarkable that some of my early results on the
thermal properties of remnants of gravitational collapse of
(quantum) matter and the early prediction of the small and
positive vacuum energy density permeating our {\it finite
Universe}~\cite{Ma96,Ma97c}, discovered by observational
astronomers and sometimes called the {\it Dark Energy}, have been
noticed by others only relatively recently. This is to say, {\it
after} the actual discovery of the {\it Dark Energy} in 1998. The
smallness of this quantity was understood as the finite size
effect in the quantum many body system~\cite{Ma96,Ma97c}, and I
quote here from~\cite{Ma96}, {\it `` The smallness of the
cosmological constant in natural Planck units is a result of an
almost perfect thermodynamical limit. This is to say that the
smallness of the cosmological constant is an effect due to an
enormous number $N$ of hypothetical \textsf{gravitational}
\textsf{atoms}. The present upper bound on the cosmological
constant $\Lambda$ allows us to draw the conclusion about the
lower bound on a number of \textsf{gravitational} \textsf{atoms}
in the observed Universe, $N\sim 10^{122}$. The large numbers'
coincidences noticed long time ago have something in it after all.
We have only one reservation to add: elementary particles are not
the same as the hypothetical \textsf{gravitational}
\textsf{atoms}, and, therefore, have little to do with the
cosmological constant. The existence of the latter must be
inferred indirectly from phenomena as it was previously done for
atoms and elementary particles. We find an unusual coordination
between the gravitational atomistic aspect of physical reality in
the regime usually called infrared, or large distance scale, and
the regime usually called ultraviolet, or short distance scale.''}

The old cosmological constant problem is a man-made problem
because the vacuum energy density has nothing to do with the
quartic divergences (zero point energies) in the interacting
relativistic quantum field theories (i.e. in the Standard Model of
elementary particles). The actual value of the vacuum energy
density of the vacuum finite size de Sitter Universe is
$\frac{E_{0}}{V}$, where $E_{0}$ is the ground state energy of $N$
\textsf{gravitational} \textsf{atoms}, which are spin-zero bosons
of mass $M\sim M_{_{Pl}}$, contained in the finite volume $V$. The
tower of excited states of such a system of $N$
\textsf{gravitational} \textsf{atoms}, which merges into the
continuum in the thermodynamic limit of $N\rightarrow\infty$,
$V\rightarrow\infty$, and $\frac{N}{V}\rightarrow n = const$, in
this limit is most conveniently described by the effective field
theory of `phonons' with an almost relativistic dispersion
relation. The usual, and incorrect, computation of the
contribution of these degrees of freedom to the vacuum energy
would involve summing over the zero point energy of `phonons'
which is at least quartically divergent. I am sure everybody would
agree that such a procedure is equivalent to the double-counting
of degrees of freedom and as such has nothing to do with the
actual value of the ground state energy of the system. I have put
it succinctly in the following phrase~\cite{Ma97c}, I quote, {\it
``...DO NOT QUANTIZE WHAT IS ALREADY QUANTIZED...''} extracted
from a footnote there~\footnote{The analogy which invites itself
quite naturally is this: in the limit $\hbar\rightarrow 0$ in
quantum statistical mechanics of ideal gases when we neglect the
fact of identical nature of quantum particles we obtain the
classical Boltzmann gas plagued with its Gibbs paradox and etc..
We should perhaps recall here the story of the now well understood
phenomenon of superfluidity in the liquid Helium II. The moral of
this very well known story, as told by R. P. Feynman, is this:
{\it DO NOT QUANTIZE WHAT IS ALREADY QUANTIZED}. This basic
observation which was found valid for superfluidity in the past is
equally valid today for quantum black holes. Landau's {\it quantum
hydrodynamics} was only formally `quantum' in view of quantum
commutators appearing in the nonrelativistic current algebra.
Feynman has discovered that the real problem with {\it quantum
hydrodynamics} was that hydrodynamical description of the
superfluid Helium II failed to take into account the Bose
statistics of Helium 4 atoms. Now, it is our opinion that we seem
to be facing the same dilemma posed by the currently fashionable
description of black holes. Incidentally, the cases for
quantization of general relativity (GRT) and its later fermionic
deformation known as `supergravity' seem to follow the same
pattern which was so well understood by Feynman in the context of
experimentally rich phenomenon of superfluidity. It seems that
general relativity, `supergravity', `superstrings' and recently
`supermembranes' were also quantized in the same way the
superfluid hydrodynamics was quantized with the well known results
. Again, the moral of the story as told by R. P. Feynman when
adopted to the present situation of quantum black holes seems to
be that we should ``give to Caesar what is Caesar's''. The Atomic
Hypothesis and Quantum Statistics rule. Gravitation is the
many-body phenomenon. It is our responsibility now to find the
physically correct Hamiltonians for systems of gravitational atoms
in the framework of the new gravitational noncommutative mechanics
.}.

This approach to the resolution of the old cosmological constant
problem has been later popularized and extended by G. E.
Volovik~\cite{VoloBook}.

\vskip .1cm A connection between the breakdown of quantum fluid
rigidity and the appearance of closed time-like curves is most
easily seen in the case of the "spinning cosmic string" or the
gravitational vortex solution of the Einstein field
equations~\cite{Ma86,Ma87}. It is the solution of these equations
only outside the vortex core; inside the core one finds CTCs. This
instance of breakdown of general relativity is also an example of
our ideal theoretical laboratory we were looking for. It turns out
that the gravimagnetic field associated with the gravitational
vortex is essentially the velocity field in the quantum fluid
surrounding the vortex core~\cite{ChapMa04}. This velocity
satisfies the Onsager-Feynman quantization condition, which
actually amounts to the Dirac-like quantization condition for the
mass-energy. The quantum fluid velocity becomes comparable to the
velocity of sound (the speed of gravity or velocity of light) when
the distance from the axis of the vortex is close to the quantum
coherence length in the quantum fluid. Therefore quantum fluid
rigidity and hydrodynamics (general relativity) breaks down as one
enters the core of the vortex. This is closely related to the
Landau criterion for the breakdown of superfluidity. Remarkably,
this breakdown of a classical description of the quantum fluid
seems to be closely related to the breakdown of causality in
classical General Relativity associated with the formation of
closed time-like curves~\cite{ChapMa04}.

\vskip .1cm The spectacular topological effect of scattering of
matter on the gravitational vortex has been
found~\cite{Ma87,Ma95}. It was found that for very slow particles
the scattering amplitude vanishes precisely for quantized values
of the mass-energy of those particles. I have concluded that if
such gravitational vortices do exist, and if they have fixed
vorticity, then there must exist a quantum of mass. Remarkably one
finds here the famous Onsager-Feynman relation which expresses the
quantum of vorticity in superfluid He4 in terms of the Planck
constant and the mass of a He4 atom. The single parameter which
appears in the gravitational vortex metric~\cite{Ma86,Ma87} has a
very simple meaning: it is the vorticity of a vortex in some kind
of isotropic quantum fluid. The Onsager-Feynman type of
quantization condition can be read in the reverse order. From the
quantization of vorticity one infers the quantization of mass in
the quantum fluid. This is the way one can see the presence of He4
atoms in the superfluid phase. It follows then that there must
also exist atoms in the normal phases of He4. The remarkable
feature of mass-energy quantization implied by the possible
existence of my gravitational vortices (`spinning cosmic strings')
may also be regarded as the harbinger of the microscopic theory of
gravitation.

\vskip .1cm The hypothesis that all gravitational phenomena, and
the recently discovered Dark Energy corresponding to the
cosmological vacuum energy density, Cosmic Microwave Background
Radiation (CMBR), primordial density fluctuations, and the most
remarkable prediction of the closely related phenomenon of global
rotation of our finite Universe, should follow from the existence
of the underlying constituent heavy particles is a bold one. It
must be supported by the line of investigation that is independent
from the arguments based on the peculiar properties of
gravitational vortices. The phenomenon of gravitational collapse
is an example of such an opening we were looking for.

\vskip .1cm It is the inconsistency of the fundamental physical
principles with the behavior of the time variable on an
analytically extended classical space-times of black holes and
cosmological models that leads to all the paradoxes of quantum
theory applied to black holes and cosmological models of our
Universe, including the so-called Hawking effect of black hole
radiance and the Bekenstein black hole thermodynamics. While this
point of view does not seem to be universally accepted yet it is
clear that accepting the possibility of the disappearance of
quantum matter past the event horizon during the process of
gravitational collapse and the failure of the existence of the
quantum ground state of gravitating matter is not an option.
Indeed, the Bekenstein-Hawking proposal for the black hole
entropy, which essentially corresponds to a classical
high-temperature thermodynamical regime, does not even seem to be
compatible with the basic statistical physics one learns from the
introductory university textbooks (Landau and Lifshitz, Kittel)
because it leads to the occurrence of negative heat capacities. A
quantum many body system is by necessity characterized by the
positive heat capacity, the property intimately related to the
unitary evolution of states of matter in quantum mechanics. {\em
The most natural resolution of the paradoxes related to black hole
thermodynamics is the observation that the Bekenstein-Hawking
proposal cannot be correct}. I have made this elementary
observation a long time ago~\cite{Ma96}. This came about as the
result of critical analysis of the above proposal along the lines
which will become easily understood by following the line of
thought described below.

\vskip .1cm One cannot underestimate the importance of the crucial
role scientific analogies play in the process of creative
thinking. The situation here should be familiar to all physicists.
In the case of black hole thermodynamics the analogy that invites
itself somewhat insistently is the analogy between the
Rayleigh-Jeans description of the cavity black body radiation and
the Bekenstein-Hawking proposal for the thermodynamics of
classical black holes. The reason being simply that each describes
the high temperature behavior of a corresponding system. The role
the Maxwell equations play in the derivation of the Rayleigh-Jeans
law in the former case is analogous to that played by the Einstein
equations in the latter. Exploring this obvious analogy I followed
it to the final conclusion but here I was helped by my earlier
work on gravitational vortices mentioned above.

\vskip .1cm Once one accepts the possibility that the pathological
thermodynamical properties of classical black holes cannot be
described by a quantum mechanical system there is no other option
left but to find a quantum system whose thermal properties at high
temperatures would correspond to the Bekenstein-Hawking
thermodynamics of the classical black hole once a certain improper
order of limits is taken. The paradoxes of BH thermodynamics
should be then seen in the completely new light. Indeed, we
encounter here an instance of non-commuting limits being taken.

\vskip .1cm One is therefore led to look for the correct
thermodynamics of cold ultra-compact gravitating systems which is
compatible with quantum mechanical principles. It is also prudent
to be careful how one takes two possible limits here because they
do not commute. The simplest way to proceed is to observe that the
heat capacity of a system is intrinsically connected with the
statistical fluctuations in the observable mass-energy of this
system. One computes the statistical fluctuations in the
mass-energy of a black hole which follow from the heat capacity
suggested by the Bekenstein-Hawking proposal and amends it in the
proper way to make them non-negative~\cite{Ma96,Ma97c}. The
physical picture of the final state of gravitational collapse of
matter, a quasi black hole (QBH), which has emerged from these
investigations is that of a cold and compact object with a surface
layer separating two regions characterized by different values of
vacuum energy density whose thermodynamics is dominated by the
presence of gapless low energy collective excitations that are not
unlike phonons in quantum
fluids~\cite{Ma96,Ma97c,MaMot98,MaMot01,MaMot04}. These are
collective excitations in a medium consisting of heavy
gravitational constituents~\cite{Ma96,Ma97c}.

\vskip .1cm The corollary of this line of investigation was the
conclusion that there should exist other phenomena where the
atomic nature of gravitation should demonstrate itself directly. I
have immediately applied this idea (in the same 1996 paper) to the
problem of the largest cold gravitating system, the finite
Universe itself~\cite{Ma96,Ma97c}. The result was the prediction
of a small, positive vacuum energy density, now known, after its
experimental discovery in 1998, as the Dark Energy. The smallness
of this quantity was understood as the finite size effect in the
quantum many body system~\cite{Ma96,Ma97c}.

\vskip .1cm If there indeed exist gravitational constituents then
one should explore the implications of this idea to the full
extent. New phenomena can be predicted on the basis of a simple
hypothesis. I have made a prediction to the effect that if the
finite Universe is rotating, which is a natural thing to
contemplate because rotation is common in nature, then this effect
should be seen as an asymmetry in the CMB spectrum for some broad
range of multipoles. This asymmetry should be most prominent for
the quadrupole and the octupole, which is to say at the largest
distance scales.

After all if our finite Universe rotates then we should be able to
detect the presence of an axis of rotation and the resulting
asymmetries. The importance of the presence of quantized vortices
in a rotating finite Universe cannot be underestimated. They make
the whole proposal work once one gets to the technical details.

\vskip .1cm \noindent What motivated me to think about the
possibility of a rotating Universe?

\vskip .1cm The reason for this were three facts of experience.
One was that galaxies have angular momentum that seems to be
related to their mass in the specific way. The second was the
problem of the origin of primordial density fluctuations and the
third was that there seems to exist a residual (Birch) rotation of
polarization of electromagnetic waves coming from distant radio
sources, although the latter continues to be considered
controversial.

\vskip .1cm There was only a small step to make which I have
already taken in the 90s. The natural idea that gravitation is the
atomic phenomenon which should be best understood as the quantum
many body problem requires consideration of the second quantized
Schrodinger equation for a system of heavy bosons with its
characteristic property which is the presence of the universal
time and the quantum entanglement (the EPR property of quantum
many body wave functions---the non-factorability of the many-body
wave function into the product of single particle `orbitals'). One
could then start to compute things on the basis of a rather simple
idea I have described above.

\vskip .2cm \noindent It appears that gravitation has something to
do with the Einstein condensation of massive bosons after all.

\end{document}